# Assessment of Sustainable Funding Impact by Exploiting Research Performance Indicators and Semantic Techniques

Muhammad Umar[a] and Saeed-Ul Hassan[a]

[a] Department of Computer Science, Information Technology University, 346-B, Ferozepur Road, Lahore, Pakistan

E-mail address: mu233@itu.edu.pk, saeed-ul-hassan@itu.edu.pk

**Abstract:** This paper deploys bibliometric indices and semantic techniques for understanding to what extent research grants are likely to impact publications, research direction, and co-authorship rate of principal investigators. The novelty of this paper lies within the fact that it includes semantic analysis in the research funding evaluation process in order to effectively study short-term and long-term funding impact in terms of publication outputs. Our dataset consists of researchers that receive research grants from the National ICT Research and Development funding program of Pakistan. We show a number of interesting case studies to conclude that bibliometric-based quantitative assessment combined with semantics can lead to building better sustainable pathways to deploy evaluation frameworks for research funding effectively.

**Keywords:** Research Evaluation; Research Grants; Sustainable Impact; ICT R&D Fund; Smart City

## 1 Introduction

The research and development (R&D) is the backbone of the globally competitive knowledge-driven economies. The investment in R&D is extremely important for the global economies to develop new products and services that drive growth, create jobs, and improve the sustainable smart city growth [1-3]. However, scientific research is becoming very expensive due to its interdependence of each branch with other disciplines [4-5]. Due to ever increasing cost of competitive research many countries are primarily finding a solution through the joint funding of laboratories and projects for sustainable outcomes [6-9] e.g. CERN[1] is the largest particle physics laboratory in the world which is operated and funded by countries in EU.

---

1 https://home.cern

While the developed world can afford the luxury of many research institutes and centers, the situation in developing world is alarming. According to Dehmer et al [10], with respect to the global R&D expenditures the developed countries are participating a lot more in research as compared to developing countries. The Unites States is currently the global leader in R&D spending having almost one third portion of global R&D spending, followed by China and Japan; however, if the unions of countries are also considered as combined units of spending, the EU then becomes second followed by China as third major contributor to global research [11].

Recently, knowledge has become crucially important for global economies, since many economies have become knowledge base [12, 13]. Many countries are striving for their economies to be knowledge based and so spending a vital part of their GDP on R&D. Israel, Finland, and Qatar are tiny (i.e. area and population wise) countries but spending higher percentage of their GDP on research in contrast to some other big countries and economies (e.g. Brazil and India). If we look into top global economies' current rates of growth and investment in R&D, China's total funding of R&D is expected to surpass that of the U.S. by about 2022 which indicates higher growth in R&D spending by China as compared to U.S [14]. The question remains that how the success and impact of R&D activities are measured? So far, publishing technical papers with maximum citations, patents, and new product introductions are considered as a key measure of success.

Among the developing nations, the research landscape of Pakistan has been very impressive since last decade and show continuous growth [15]. Research trends are improving and initiatives have been taken to promote innovation through research projects and ideas. Based on the research output statistics, Pakistan is being expected to be second fastest growing country in research output just after Malaysia [16]. Due to scarce resources (e.g. funds) in Pakistan and ever increasing cost of carrying out cutting edge research, it is hence necessary to make best use of funds and scientific manpower to achieve high scholarly impact. It thus becomes important to develop a formal understanding of the key factors that influence the relationship between funding, research output and impact, and to develop methods to meaningfully measure the impact of research funding [17].

In this paper we study the relationship between research funding an, research output and impact with the objective of developing a formal analytical framework. We employ metrics for evaluating various dimensions of research output and impact for the accurate measurement of national research productivity. An important aspect of this study is the examination of the longer-term impact of funding beyond the grant period - which provides information about funding impact that goes beyond the time embodied in the final reports commonly submitted at the end of a grant. We examine Pakistani researchers that receive funds from National ICT R&D fund program – a flagship national research-funding program that aims "to transform Pakistan's economy into knowledge based economy by promoting efficient, sustainable and effective ICT initiatives through synergic development of industrial and academic resources" [2]. The followings are the object of this study:

- Analysis of performance indicators relating funding to research productivity and impact.
- Studying research productivity per unit of funding at organization level and at individual researcher level.
- Long-term sustainable impact of research funding on research activity beyond the duration of the grant by deploying semantic analysis on the scientific literature published by the scholars.

The novelty of this paper lies in the fact that a lot of work has been done by deploying various bibliometric indicators to better describe the relationship between researcher, research, funding, and impact of research funds; but no significant work has been done so far to include semantics during research funding evaluation process and to effectively study the grant allocation process for the qualitative and quantitative measure of sustainable scientific impact. The rest of the paper has been structured as follows: Section 2 includes previous work related to methodologies of impact assessment of funding, data and text mining, importance of semantics in revealing trends, and some discussion on the challenges in assessing impacts of funding. Section 3 describes the datasets, data collection procedures and analytical techniques to measure similarity and closeness among documents. Section 4 includes all the major analysis and measure carried out during this study. Section 5 presents a case study a case study to illustrate the use of our deployed measures to study the impact of ICT R&D fund project. Finally, Section 6 presents conclusions and makes some recommendations for future work.

---

[2] http://www.ictrdf.org.pk/

## 2 Literature Review

This section reviews the related literature from the following two perspectives: In the first half we present a brief review on the performance indicators relating funding to research productivity, in the second part, we present a brief review on related semantic and text mining techniques used to study the long-term impact of research funding on the scientific literature published by the scholars.

### 2.1 Review on research performance indicators

The R&D funding programs play vital role in public research policy and these schemes leading towards improving quality research output. The sole purpose located beneath these programs is to make sure that competitive grants are helping to enhance research performance and to get the most of it. Studying the impacts of such funding schemes have become more common in such a way that it makes funding authorities more curious and keener to make sure that these grants have intended positive impacts on research performance and scientific quality [18].

Several studies examine the relationship between funding and research output at the level of the university or department; some examine international scientific knowledge flows and scholarly impact of countries and institutions [19, 20]. Other studies we came across examine the funding-output link at the level of the individual researcher and found small positive effects [21, 22]. A few studies provide impact assessment for progressing research and some worked on assessing economic impact of R&D [23, 24]. Results of these studies show higher increases in the number of publications for grant recipients than for rejected applicants, while increases in mean normalized citation rates were not significantly higher for the successful applicants [25, 26]. However, it should also be noted that these increases in productivity also include greater productivity of highly cited papers [18].

A few studies have focused on inspecting the connection between researchers past performance and the peer review of grant proposals, and found that the application candidates have a tendency to have better track records in the form of higher citation scores than non-applicants and rejected candidates [27, 28, 29]. Similarly, Chilean research funding found a noteworthy impact on number of publications (i.e. higher research output for successful candidates in contrast to rejected candidates), but no significant impact found on citations [30]. Furthermore, Jacob & Lefgren [31] carried out study for NIH postdoc grants impacts. They found positive impact on research output for granted applicants i.e. the number of publications increased with the reception of grant, also increase in probability of crossing a citation threshold has noted, but total number of citations got no impact for successful and unsuccessful applicants. Melin & Danell [32] showed that young investigators in Sweden whose applications were selected for 6-years grant didn't impact the number of publications as expected (no improvement for awarded applicants in research output as compared to rejected ones), however positive impact has found in terms of international co-authorship, which helped the research groups for securing furthers funding.

Recently, Langfeldt et. al [33] study the impact of Norway's FRIPRO and Denmark's DCIR funding agencies and found significant impact of the funding schemes in both countries. The higher increase in research output and highly cited paper had seen for successful applicants in contrast to rejected candidates. However, did not found any notable impact on average citations for successful and rejected applicants, concluding no impact on importance of research.

## 2.2 Review on text mining and semantics

Data Mining and Text Mining approaches can be utilized to find patterns from structured data [34]. These techniques can be applied on unstructured scholarly big data corpus to convert it into structured data in order to reveal valuable insights. Text mining can be also used to discover and extract patterns and trends in scholarly data by deploying Natural Language Processing (NLP)[35]. The NLP is a technique that is widely used to extract information from textual data. NLP is particularly effective in the extraction of predefined patterns or existing information that can help in finding and exploring trends from any database [36].

An investigation on document similarity found that commonly used similarity techniques such as the cosine and Jaccard treat the words as independent entities from one another, which is however unrealistic as words in documents combine together to deliver proper context. Words in any document are interrelated to form meaningful structures and to develop ideas. An alternate is suggested to use concepts instead of words to extract the topics of documents by resolving redundancies (i.e. synonymy) and ambiguities (i.e. polysemy) in words [37].

Another study also describes the same weaknesses (i.e. treating words as independent entities) of BOW (Bag of words) approach used in common similarity measures and suggests an alternative to overcome the situation and to get improved accuracy. BOW typically represents the text in vector space model and does not consider redundancies (i.e. synonymy) and ambiguities (i.e. polysemy) problems and ignores semantic relatedness among words. To overcome the shortages of BOW approach, the suggested alternative is to embed WebNet based semantic relatedness measure for pairs of words, into a semantic kernel. This measure incorporates the TF-IDF weighting scheme, thus semantic and statistical information is combined from text that provide improved classification accuracy [38, 39].

In this study, we use vector-space model to represent text obtained from the scientific document. Thus, the text is represented by the vectors of terms extracted from the documents, associated weights are then assigned to define the importance of these terms in respective document and collection of documents under consideration. The weights (see Equation 1) of terms are normally calculated using TF-IDF method, in which the weight of a term is determined by two factors: how often the term j occurs in the document *i*, i.e. the term frequency $tf_{i,j}$ and how often it occurs in the whole collection of documents i.e. the document frequency $df_j$ [40].

$$w_{i,j} = tf_{i,j} \times \log\left(\frac{N}{df_i}\right) \qquad (1)$$

After term weights are determined, ranking function is used to measure similarity between the query and document vectors. Cosine measure is commonly used similarity measure which is used to calculate the angle between the query vector and documents vector when they are represented in a V-dimensional Euclidean space, where V is the vocabulary size [40], see Equation 2.

$$sim(Q, D_i) = \frac{\sum_{j=1}^{v} w_{Q,j} \times w_{i,j}}{\sqrt{\sum_{j=1}^{V} w_{Q,j}^2 \times \sum_{j=1}^{V} w_{i,j}^2}} \qquad (2)$$

We presented a brief review on the related literature on the performance indicators relating funding to research productivity. In addition, we also discussed techniques to semantically analyze the data using text mining and clustering techniques. While performance indicators employed in this study include publications/unit of funding, research output growth for PIs and PIOs, growth in researcher co-authorship rate, the semantic analysis are carried out to study the long-term impact of research funding on the scientific literature published by the scholars for assessing research focus or scholars and long term funding impact. Combining both statistical bibliometric and semantic analysis for performance/impact measure lead to better understand relationship between research output and research funds.

## 3  Data and Method

Text mining techniques including information retrieval and data mining have been used to carry out the analysis. In addition, web scraping technique has been used to automate data retrieval process of ICT R&D funded projects data and wrappers/parsers have been developed to extract formatted data from Scopus based All Pakistani research publications from CSV dataset. Furthermore, text processing, vector space modeling and similarity/distance measures have been employed to find out similarity among documents. Bibliometric analysis has been carried out to assess impact between funding, proposals, researchers, organizations and publications.

### 3.1  Dataset

To carry out the analysis, two types of datasets were needed i.e. funding agencies funded projects data and researchers' publications output data. First dataset include the information of funded projects by the ICT R&D funding agency of Pakistan available at their website[3]. The second dataset belongs to Pakistani researchers publications data downloaded from Scopus. The datasets include information about Principal Investigators (PIs), Principal Investigators Organizations (PIOs), abstract or summary of proposal and output, funding amount, duration and year of grant and publications.

The funding data, obtained from ICT R&D public records, made limited to only closed funded projects during 2007 to 2013. These are the projects that were funded and are completed successfully and has some output available for long term impact evaluation. We did not include information of ongoing funded projects in our dataset because that might mislead results, as it doesn't include required attributes for analysis.

---

[3] http://www.ictrdf.org.pk/

All Pakistani researchers publications dataset was extracted from Scopus data. It was then preprocessed to exclude unnecessary attributes, and filtered to include only publications during 2005-2013 for the sake of simplicity; and was kept in CSV format for further processing and analysis. This dataset includes information of more than sixty thousand publications (output of around forty thousand unique researchers that belong to more than two hundred distinct organizations). The followings are the statistics of data used in this study: publications= 61,421; researchers=42,376, organizations=213; funded projects= 17, funded researchers = 23 and funded organizations =10.

## 3.2 Methodology

To evaluate performance of overall research funding program, funding agencies ought different impact evaluations of their funding schemes to see its effects on to research performance and scientific quality. Our approach to evaluate impacts of funding has been divided into four different stages. These stages are: 1) grant allocation, 2) research output, 3) research collaboration, and 4) long term research impacts [41].

During *grant allocation* stage, the funding recipients are characterized, and analyses of funding allocation patterns are carried out as a method for assessing impacts of research activities. This quantitative data can then further be utilized in other comparisons and analyses e.g. comparisons between funding recipients and rejected candidate etc. In second stage of *research output* evaluation; analysis of scientific output related to funding is carried out i.e. calculate the number of publications produced, as well as by identifying factors that influence the output volume e.g. funding duration, research team etc. The main focus of the analysis is to find out opportunities and outcome brought about by research funding (see Equation 3).

$$productivity_{pub} = avg(publications)_{ag} - avg(publications)_{bg} \quad (3)$$
$$where\ pub = publications,\quad ag = after\ grant,\quad and\quad bg = before\ grant$$

The third stage includes impact of research funding on research users; which is not linear and hard to identify and quantify; so *research collaboration* is assumed as a proxy measure for impacts evaluation because it has positive impact on the performance of the innovation system. The analysis includes research collaboration impacts evaluation of funding by examining increased or decreased average co-authorship rate after funding (see Equation 4).

$$impact_{co\_auth} = avg(co\_auth\ rate)_{ag} - avg(co\_auth\ rate)_{bg} \qquad (4)$$
$$where\ co\_auth = co\ authorship,\quad ag = after\ grant,\quad and\quad bg = before\ grant$$

The final stage deals with *long-term impact* evaluation; analysis focusing directly on the impacts on research users is carried out. The assessment of such type of impacts is very challenging because it includes employing both qualitative and quantitative approaches to measure aggregated impacts. These impacts include evaluation of research direction focus, productivity and other indicators of individual researcher with respect to funding (see Equation 5).

$$impact_{rf} = avg(own\ similarity)_{ag} - avg(own\ similarity)_{bg} \qquad (5)$$
$$where\quad rf = research\ focus,\quad ag = after\ grant\ ,and\quad bg = before\ grant$$

Also during this phase most appropriate candidates' suggestion is made against any research proposal based upon candidate's publications similarity with respect to research proposal summary. This analysis helps to evaluate researcher's expertise in the area he/she has applied for funding compared with all the researchers in the dataset. The outcomes of this analysis can also be used to find appropriate reviewers working in related areas. The candidates with higher/maximum publications similarity with research proposal are shown as Equation 6.

$$candidate\ suggestion = max\left(avg_{i=1}^{m}\left(sim_{j=1}^{n}(P, Q_{i,j})\right)\right) \qquad (6)$$
$$where\ P = research\ proposal,\quad Qij = ith\ researcher's\ jth\ publication\ abstract,$$
$$m = total\ researchers\ ,\quad and\quad n = number\ of\ publications\ of\ mth\ researcher$$

The applied document similarity algorithm in Equation 6 has broken into three major parts; pre-processing of documents, document's text to vector space model, and documents similarity measure. For this process, documents are passed through pre-processing phase which involved first tokenizing the documents, then these tokens are transformed to same casing process e.g.

lower casing, and finally stop-word removal filters and stemming processes are applied. After pre-processing of documents, it is then mapped to a vector and weights are assigned to each term on the basis of TF-IDF, which assigns weights on the basis of term importance and its occurrence in document collection. After assignments of weights, final step is to calculate similarity between documents (project or publication summary), for this purpose, cosine similarity measure has been taken into account, which is quite commonly used measure. Cosine similarity measure is used to calculate similarity between documents based on the weights of the words represented by the two document vectors as shown in the Equation 7.

$$w_{i,j} = tf_{i,j} \times \log(N/df_i) \qquad (7)$$

The result of cosine similarity is always between zero and one. While zero indicating no similarity between documents, one indicating that documents are same. The high similarity output indicates that documents are more similar and closely related to each other whereas low similarity results indicate that the documents are less similar and more different. The formula to compute cosine similarity is presented in Equation 8.

$$similarity = \cos(\theta) = \frac{A.B}{||A||\,||B||} = \frac{\sum_{i=1}^{n} A_i \times B_i}{\sqrt{\sum_{i=1}^{n}(A_i)^2 \times \sum_{i=1}^{n}(B_i)^2}} \qquad (8)$$

## 4 Results and Discussion

In this section we deploy bibliometric performance indicators and semantics to understand PIs' research focus, diversity, productivity and co-authorship impact w.r.t funding. The purpose of these analyses is to measure the research productivity of PI after the grant begin received, co-authorship rate (research collaboration) and research focus related to theme the received grant. The data is divided into two periods, one before (excluding application year) and one after (from funding year onwards) the funding decision. The differences between the two periods are analyzed on the basis of the following indicators: a) Researchers' average number of publications per year; b) Researchers' average co-authorship rate per year; c) Researchers' research focus w.r.t. funded theme per year and d) Research organizations' funding and productivity statistics.

First of all, funding agency's data is selected for carrying out analysis. The purpose of these analyses is to see the trends in funding data, to which the funds are allotted, agency's yearly grants, yearly funded projects, number of researcher per project, average funding per project, researcher wise grants, and organization wise grants, and years in which maximum and minimum grants are allocated etc. Funding dataset includes researcher names, proposal summary, and grant information, year of grant, and status of project i.e. closed or in progress. This study is carried out for only funded projects which are in closed status because one intention of this study is to investigate the long-term impact of funding.

Secondly, all Pakistani research publications data is utilized for analysis to see the behavior of data i.e. researcher's total and yearly publications, organizations total and yearly publications, publications average co-authorship rate etc. These analyses when combined with funding agency's data become very helpful in understanding flow of funding as well as researchers and organizations research trends with respect to grants. Funding impacts on research organizations have also assessed during this study but are made limited to only productivity analysis (increase or decrease in research output) for the sake of simplicity and limitations of scope (future works can cover other indicators including quality of research etc.). The Figure 1 lists all the funded organizations along with the information of total assigned projects by the funding agency, total allotted funds by the agency, and the productivity (in terms of average increase in publications w.r.t grant). All organizations depict positive impacts of funding on their productivity. For this analysis funded organization data is obtained along with their names mapping (funding agency's mentioned organization names and Scopus data-set used organization names) and then publication statistics are extracted from Scopus dataset for selected organizations. Note that the funds are shown in PRK (104 PRK ~ 1 USD).

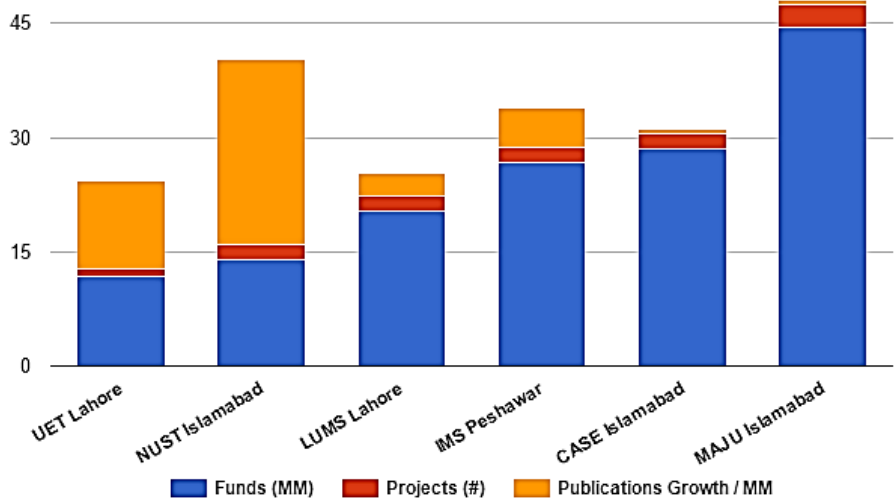

**Figure 1:** Funded organizations productivity w.r.t funding in PKR

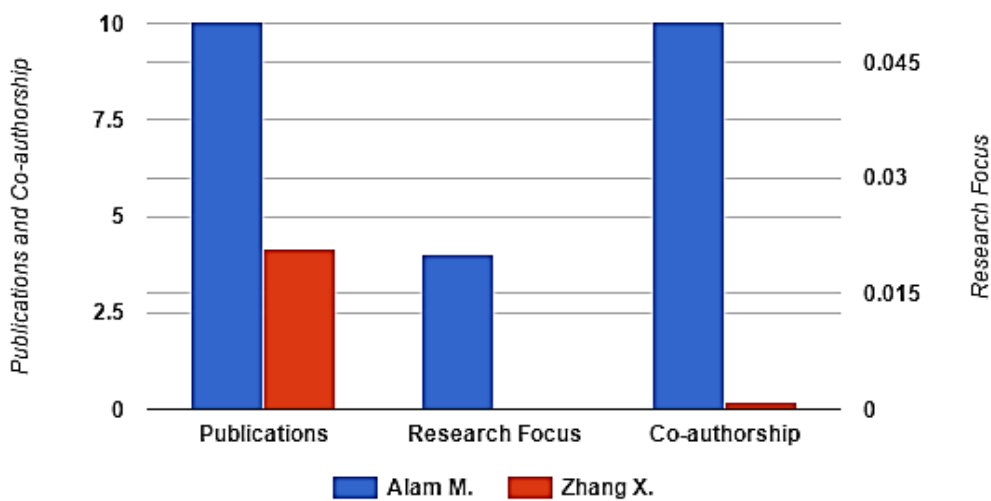

**Figure 2:** Funding impact on funded researchers' publications and co-authorship

Impacts of funding on researcher's collaboration are also evaluated on the basis of average growth (positive or negative) in co-authorship rate comparing two periods of publications i.e. prior and posterior to grant. Other semantic analysis include assessing impacts of funding on research focus and diversity of researchers. For this purpose, each researcher's publications are semantically analyzed (similarity measures are carried out among researcher's own publications) to see how much specific an individual researcher in his/her research area. The higher similarity shows more area-specific research, and less diversity and vice versa. The Figure 2 shows results of such analysis for two researchers where they are mapped along with their co-authorship rate impact (increase or decrease in average co-authorship rate after allocation of funds) and research focus impacts (positive or negative impact comparing before and after grant periods). Higher positive funding impacts on research focus and co-authorship can be seen for PI Alam M. in comparison with PI Zhang X. having no funding impact on research focus means the PI's research focus did not shift towards the theme funded was granted and very low positive impact on co-authorship.

Another analysis on funded researchers at individual level is carried out to see the researcher's yearly progress and funding impacts. The Figure 3 shows the case of a funded researcher Zaheer A. whose yearly productivity (increase or decrease in publications count) and impacts (average increase or decrease in co-authorship rate and research focus) before and after grant are mapped and funding year which is 2010 in this case is marked as green background. Positive funding impacts on publications growth and research focus are found in this case, however; research focus has observed slightly negative funding impact.

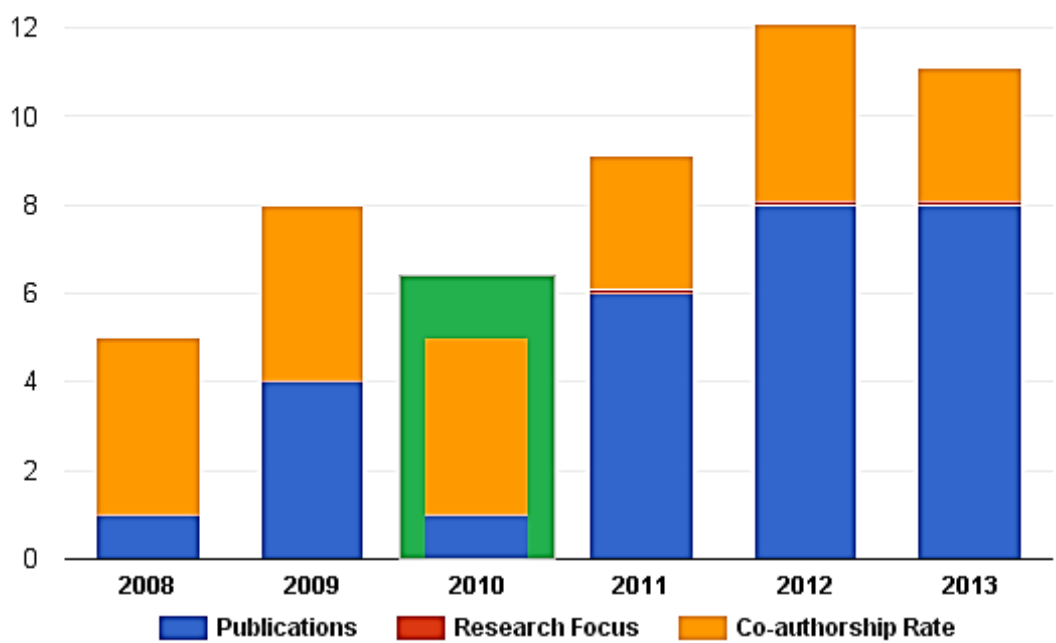

**Figure 3:** Funded researcher's (Zaheer, A.) yearly productivity funding impacts

Interesting results have seen during analysis phase. In an analysis of organizations allocated funds and research output, positive impact (more research output after allocation of funds) has been found for almost all organizations. However, some organizations have been found more productive than others in terms of higher publication rate per unit of fund. While analyzing researchers productivity and impact measure w.r.t funding, positive impact has been seen for most of the researchers i.e. higher research output rate (positive difference) found comparing two periods i.e. before and after grant. However 'No' or negative impact (equal or less research output rate after allocation of funds) of funding has also been observed in some cases. Hence, researchers' performance is on average better after the grant. In examining impact of funding on co-authorship rate of researchers, again positive impact of funding has been seen for most of the cases and negative impact for very few cases; depicting overall positive funding impact on researcher's co-authorship rate. However, mixed results has been found while measuring researcher's research focus and diversity of research, before and after grant i.e. in most of the cases no funding impact has found, positive and negative impact has found for rest of the cases; overall resulting no significant funding impact on researcher's research focus. Evaluations of the impacts and productivity analysis can be seen in Figure 4.

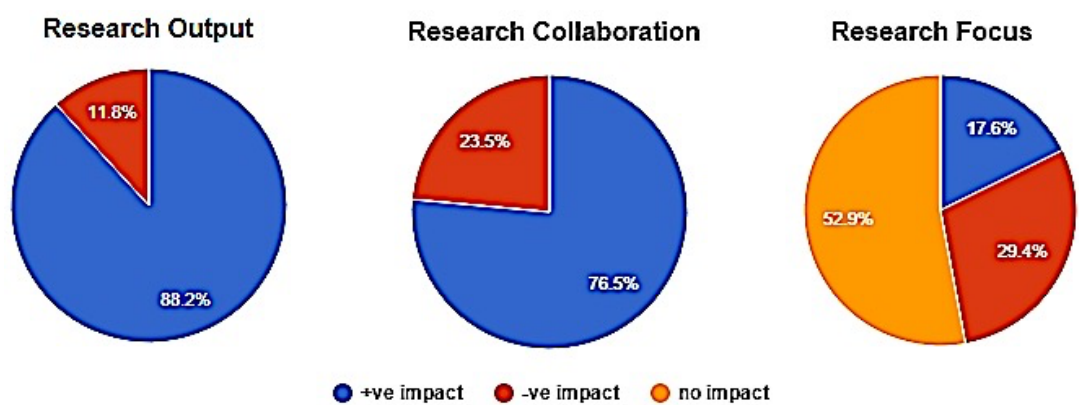

**Figure 4:** Researchers funding impacts evaluations

Suggested candidates (potential reviewer or collaborator) analysis has also been performed against each funded projects to list down best possible match candidates (based upon their higher research similarity with the project and higher publications average) to whom funding agency can coordinate with, during proposal evaluation process, funding allocation process and/or ongoing project assistance.

## 5   Case Study: Measuring impact of National ICT R&D funded project

We present a case study to illustrate the use of our deployed measures to analysis the impact of a ICT R&D funded project. For this purpose we select a project out of available seventeen closed projects funded by ICT R&D funding agency during 2007-2013. We then analyze PIs and PIO productivity and funding impacts of selected project; theses analysis are carried out on all Pakistani research output data extracted from Scopus during 2005-2013. In addition, we suggest most appropriate list of candidates who could be potential reviews or collaborators having higher research relevance with the selected project. Finally, we discuss the results of semantic and bibliometric analysis of the selected project.

**Table 1:** Descriptive information of selected funded project

| Project Title | Power Aware Video Coding for Extending Battery Life in Portable and Mobile Devices. |
|---|---|
| **PIs** | Dr. Nadeem Khan, Dr. Jahangir Ikram |
| **PIO** | Lahore University of Management Sciences (LUMS), Lahore |
| **Year** | 2009 |
| **Budget** | PKR 13.03 million (~ 130k USD) |

We select a successfully closed funded project to analyze funding impacts on researcher, organization and research work itself. The information required to carry out necessary analysis include project title, project summary, principal investigator's name, principal investigator organization, funding year, and allocated budget (see Table 1). This project was funded in 2009 and our dataset contains research information from 2005 to 2013, so long term impacts can be easily assessed.

We analyze selected organization's i.e. Lahore University of Management Sciences, Lahore (LUMS) productivity to see either selected organization has observed any positive impact of funding in terms of research output growth. Organization's average increase in research output statistics is mapped along with their assigned funded projects and grants by the ICT R&D funding agency. Positive funding impact on organization's publications growth i.e. around three publications per million PKR can be seen in Figure 5.

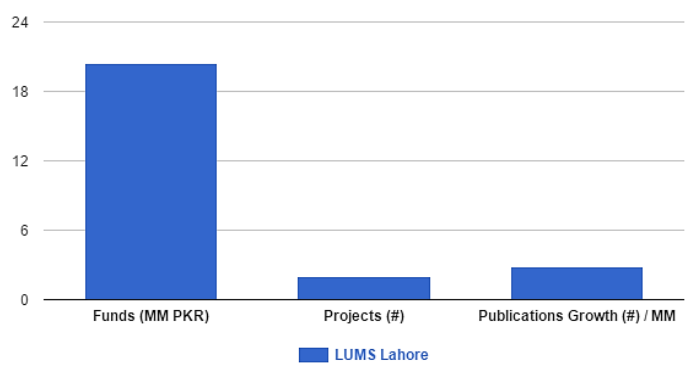

**Figure 5:** Organization's funding and productivity statistics 2005-2013

We then analyze funding impacts at individual researcher level. Principal investigators are assessed on the basis of average publications growth, increase or decrease in research focus, and positive or negative change in average co-authorship rate before and after funding. The Figure 6 depicts positive funding impact on research output and average co-authorship for both the investigators under consideration; however, no funding impact found on research focus of investigators i.e. research focus remained almost unchanged after grant allocation.

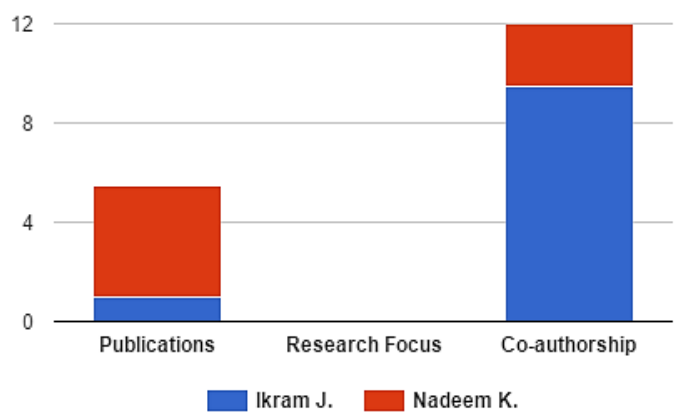

**Figure 6:** Principal investigators funding impacts statistics

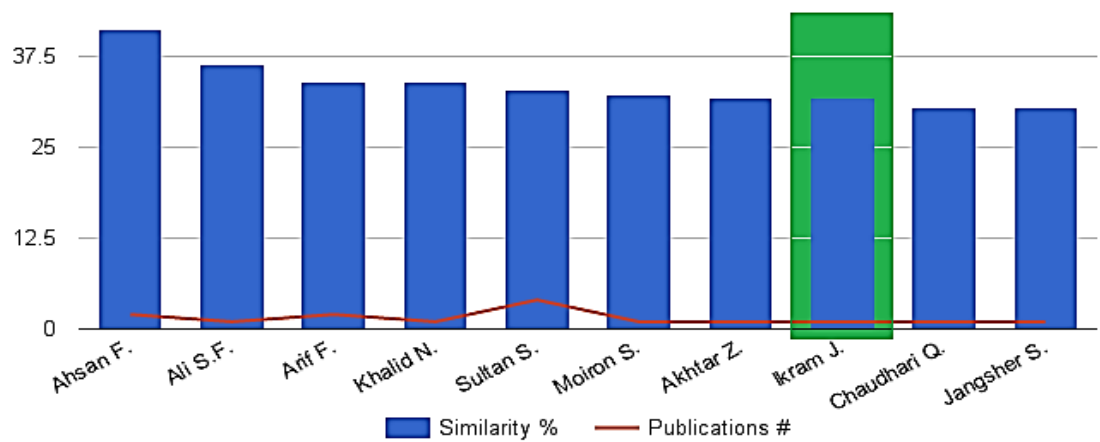

**Figure 7:** Candidates suggestions based upon research similarity

Finally, we suggest best possible match candidates for collaboration or potential reviewers against selected project by semantically analyzing executive summary. The project's executive summary is matched with each researcher's publications abstract in dataset. Finally, the researchers with similar fields are extracted and then highly similar researchers are suggested. In Figure 7, suggested researchers are mapped along with their research output statistics. We find one of the PI Ikram J. appears in the list, suggesting that the researcher has expertise in the field to execute this project.

# 6 Concluding remarks

This is a novel approach towards impact assessment that deals not only with the bibliometric analysis to evaluate impacts of funding on research but also deploys semantic analysis to make this evaluation more appropriate and accurate. This study examines the measures, period prior to funding and after - the difference between before and after the grant being received. Positive funding impact on research output (i.e. number of publications) has found for almost all funded researchers and research organizations. The approach of using semantics along with bibliometric indicators (relating funding and impacts) can be very helpful in making funding program more effective and for better impacts evaluation; it is recommended for funding agencies to use it in formal framework formation and/or proposal evaluation process.

Similarity based analysis during this study are carried out using cosine similarity measure which is very simple and efficient but treat the words independently, thus, does not include relatedness between the words, which may lead to poor results. Better and more accurate results can be achieved by embedding semantic relatedness [42-44] between words and resolving redundancies and ambiguities as examined in following studies [45-48]. Future works can also include suggesting funding agencies about currently hot research areas rest of the world is working using deep learning methods [49-53]. These research trends evaluation when combined with funding impact evaluation might lead to better framework development for funding agencies which covers the processes of area selection, proposal evaluation, ongoing research assistance, and finally long-term sustainable impact of research assessment.